\documentclass[aps,prl,reprint, showpacs, showkeys, twocolumn, superscriptaddress, notitlepage]{revtex4-1}

\usepackage{graphicx} 
\usepackage{braket}
\usepackage{color}
\usepackage{times}
\usepackage{amsmath}
\usepackage{lipsum}
\usepackage{multirow}
\usepackage{adjustbox} 
\usepackage{textcomp} 
\usepackage{soul} 
\usepackage{hyperref}

\DeclareTextSymbol{\degre}{T1}{6}
\DeclareTextSymbol{\degre}{OT1}{23}

\begin{document}

\title{On-demand semiconductor source of 780\,nm single photons with controlled temporal wave packets}

\author{Lucas B\'eguin}
\affiliation{Department of Physics, University of Basel, Klingelbergstrasse 82, CH-4056 Basel, Switzerland}

\author{Jan-Philipp Jahn}
\affiliation{Department of Physics, University of Basel, Klingelbergstrasse 82, CH-4056 Basel, Switzerland}

\author{Janik Wolters}
\affiliation{Department of Physics, University of Basel, Klingelbergstrasse 82, CH-4056 Basel, Switzerland}

\author{Marcus Reindl}
\affiliation{Institute of Semiconductor and Solid State Physics, Johannes Kepler University Linz, Altenbergerstrasse 69, A-4040 Linz, Austria}

\author{Yongheng Huo}
\affiliation{Institute of Semiconductor and Solid State Physics, Johannes Kepler University Linz, Altenbergerstrasse 69, A-4040 Linz, Austria}

\author{Rinaldo Trotta}
\affiliation{Institute of Semiconductor and Solid State Physics, Johannes Kepler University Linz, Altenbergerstrasse 69, A-4040 Linz, Austria}

\author{Armando Rastelli}
\affiliation{Institute of Semiconductor and Solid State Physics, Johannes Kepler University Linz, Altenbergerstrasse 69, A-4040 Linz, Austria}

\author{Fei Ding}
\affiliation{Institut f\"ur Festk\"orperphysik, Leibniz Universit\"at Hannover, Appelstrasse 2, D-30167 Hannover, Germany}

\author{Oliver G. Schmidt}
\affiliation{Institute for Integrative Nanosciences, IFW Dresden, Helmholtzstrasse 20, D-01069 Dresden, Germany}

\author{Philipp Treutlein}
\affiliation{Department of Physics, University of Basel, Klingelbergstrasse 82, CH-4056 Basel, Switzerland}

\author{Richard J. Warburton}
\affiliation{Department of Physics, University of Basel, Klingelbergstrasse 82, CH-4056 Basel, Switzerland}

\date{\today}

\begin{abstract}

We report on a fast, bandwidth-tunable single-photon source based on an epitaxial GaAs quantum dot. Exploiting spontaneous spin-flip Raman transitions, single photons at 780\,nm are generated on-demand with tailored temporal profiles of durations exceeding the intrinsic quantum dot lifetime by up to three orders of magnitude. Second-order correlation measurements show a low multi-photon emission probability ($g^{2}(0)\sim\,0.10-0.15$) at a generation rate up to 10\,MHz. We observe Raman photons with linewidths as low as 200\,MHz, narrow compared to the 1.1\,GHz linewidth measured in resonance fluorescence. The generation of such narrow-band single photons with controlled temporal shapes at the rubidium wavelength is a crucial step towards the development of an optimized hybrid semiconductor-atom interface.

\end{abstract}

\maketitle

\section{Introduction}

The distribution of quantum states and entanglement between remote systems within a quantum network~\cite{Kimble2008} enables a vast range of technological breakthroughs from secure communications~\cite{Lo2014} to computational speed-up~\cite{Nielsen2000} and quantum-enhanced global sensing~\cite{Komar2014}. In this framework, single photon sources are essential resources that allow matter qubits at stationary network nodes to be interconnected~\cite{Grangier2004,Eisaman2011,Sangouard2012}. For most of these applications, controlling the spectral and temporal properties of the single photons is a crucial requirement. Indeed, the performance of quantum protocols based on two- or single-photon interference critically depends on the degree of coherence of the individual photons: the coherence limits the achievable coalescence contrast in two-photon interference experiments~\cite{Mandel1991}. Single-photon wave packets should therefore be generated in a well-defined spatio-temporal mode with a Fourier-transform-limited spectrum. The ability to tailor the photons' carrier frequencies, spectral widths and temporal profiles is essential to ensure efficient coupling between remote heterogeneous systems~\cite{Raymer2012}. 

In particular, control over the temporal profile, the waveform, of the single photons is important for a number of reasons. First, ``long" single photons with narrow spectra are required for an efficient interaction with media featuring sharp absorption lines such as atomic species or solid-state color centers. Secondly, protocols for long-distance entanglement distribution require path length differences stabilized to within the temporal ``length" of the single photon wave packets~\cite{DeRiedmatten2004}, and the use of long photons thus relaxes these requirements. Finally, fine control of the temporal profile enables the coupling efficiency between single photons and atoms~\cite{Gorshkov2007a,Aljunid2013} or between single photons and optical cavities~\cite{Liu2014} to be optimized. Numerous approaches to generate single photons with tunable spectro-temporal properties have been investigated using cavity-enhanced spontaneous parametric down conversion~\cite{Bao2008,Haase2009,Fekete2013,Rambach2016}, single atoms~\cite{McKeever2004,Hijlkema2007,Nisbet2011} or ions~\cite{keller2004,Barros2009} in a cavity, hot~\cite{Eisaman2004,Willis2011,Ding2012,Shu2016} and cold~\cite{Laurat2006,Matsukevich2006,Thompson2006,Du2008,Bao2012,Srivathsan2013,Zhao2015,Farrera2016} atomic ensembles, trapped ions in free space~\cite{Kurz2013}, and quantum dots~\cite{Fernandez2009,Rakher2011,Matthiesen2013,He2013a,Sweeney2014,Lee2017}. 

Among all single photon emitting devices, semiconductor quantum dots (QDs) embedded in dedicated photonic nanostructures are highly promising single photon sources. QD sources combine simultaneously large photon extraction, high brightness and near-perfect levels of purity and indistinguishability~\cite{Ding2016,Somaschi2016}, all in a fast and robust device. These properties are not shared by any other source. An exciton, an electron-hole pair, mimics a two-level system in these devices. However, exciton recombination takes just a few hundred picoseconds such that QDs usually generate single photons with GHz linewidths. This linewidth far exceeds the bandwidth of prototypical single photon memories. A specific and important example is an ensemble of atoms which have excellent properties for a photon memory~\cite{Bussieres2013} but only in a narrow bandwidth, typically $\sim 10$\,MHz. Interfacing GHz-bandwidth single QD photons with atomic memories is therefore highly inefficient on account of the bandwidth mismatch.

Finding a way to control the spectro-temporal properties of QD photons represents a key challenge. In this direction, several methods have been investigated. A first temporal shaping demonstration implemented fast electro-optic amplitude modulation synchronized with the photon generation to temporally filter preselected profiles from exponentially decaying envelopes~\cite{Rakher2011}. Although this method can help to improve the degree of indistinguishability of a noisy source, it works by introducing losses that significantly reduce the effective brightness. Another approach exploited weak resonant excitation to generate highly coherent, indistinguishable photons with tailored waveforms via Rayleigh scattering~\cite{Matthiesen2013}. However, such a method prohibits the generation of single photon wave packets of durations longer than the QD exciton lifetime and therefore does not address the bandwidth mismatch with atomic memories.

The two-level exciton offers a too restrictive set of possibilities. Inspired by experiments on trapped ions~\cite{Almendros2009,Muller2017}, a much more powerful approach is to create a three-level system, specifically a $\Lambda$-system, by trapping a single electron or hole in the QD. A $\Lambda$-system is created on application of a magnetic field. The main idea is to generate a single photon with tailored waveform by driving the spin from one spin state to the other, a Raman process. While Raman scattering from QD $\Lambda$-systems is established~\cite{Fernandez2009,He2013a}, forming in fact the basis for recent demonstrations of remote spin entanglements~\cite{Delteil2016,Stockill2017}, the creation of on-demand single-photons with user-defined temporal profiles is not. 

Here, we demonstrate high-rate, on-demand generation of single photons with tailored temporal wave packets from a QD. The QD is spectrally matched to the rubidium D2 line. Such a versatile single photon source opens up important applications in heterogeneous quantum networking, combining tailored single photons with broadband atomic quantum memories~\cite{Saunders2016,Wolters2017}.

\section{Scheme}

\begin{figure*}[ht]
\includegraphics[width=150mm]{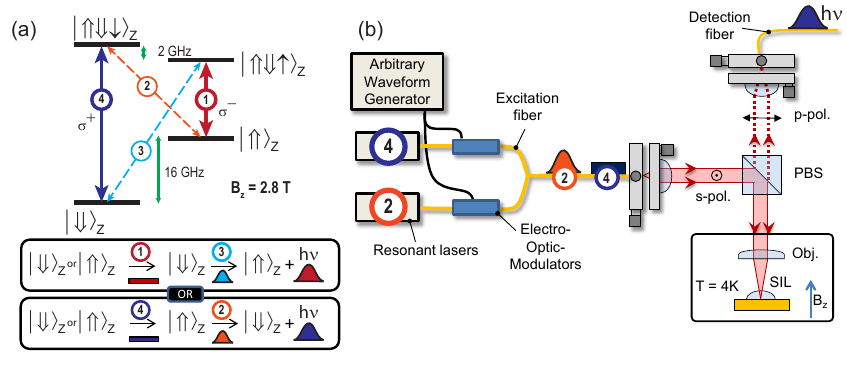}
\caption{(a) Reduced energy level diagram of a quantum dot charged with a single hole subject to a magnetic field in the Faraday geometry. The inset illustrates the two-step sequence used for generating a ``red" or ``blue" single Raman photon with controlled temporal waveform. (b) Polarization-based dark-field microscope with tailored excitation pulses. }
\label{Fig1}
\end{figure*}

We consider a QD charged with a single hole. The ground states correspond to the two hole spin states; the excited states to the two trion states. In a magnetic field along the growth direction, the hole spin ground states ($\ket{\Uparrow}_z$ and $\ket{\Downarrow}_z$) and the trion excited states ($\ket{\Uparrow \Downarrow \uparrow}_z$ and $\ket{\Downarrow \Uparrow \downarrow}_z$) are split in energy according to the out-of-plane $g$-factors $g_h$ and $g_e$, respectively (Fig.~\ref{Fig1}(a)). For a pure heavy-hole state, selection rules dictate that only the ``vertical" spin-preserving transitions (\raisebox{.5pt}{\textcircled{\raisebox{-.9pt} {1}}}  and \raisebox{.5pt}{\textcircled{\raisebox{-.9pt} {4}}}) are allowed with orthogonal circular polarization ($\sigma^-$ and $\sigma^+$, respectively). In practice, the ``diagonal" spin-flipping transitions (\raisebox{.5pt}{\textcircled{\raisebox{-.9pt} {2}}} and \raisebox{.5pt}{\textcircled{\raisebox{-.9pt} {3}}}) are also weakly allowed by heavy hole-light hole mixing or by the hyperfine interaction (the nuclear spins induce a slight tilt of the quantization axis)~\cite{Warburton2013}. This means that each trion state possesses two spontaneous decay channels, one fast, the other slow. This is described as a $\Lambda$-system with a very asymmetric branching ratio, $\gamma / (\Gamma+\gamma) \ll 1 $, where $\Gamma$ and $\gamma$ are the ``allowed" and ``forbidden" spontaneous decay rates. As a result, optical spin pumping is achieved by resonantly driving the strong spin-preserving transitions until the trion spontaneously decays via the weak spin-flipping transitions~\cite{Atature2006, Gerardot2008}. Once the QD spin state has been initialized, a single photon can be generated by driving the weak ``diagonal" spin-flipping transition of the $\Lambda$-system. A single photon is generated on driving the spin from one spin state to the other. This is a Raman process. The asymmetric branching ratio ensures that the purity of the photon scattered in the spontaneous Raman process is not limited by an otherwise broad emission time distribution~\cite{Muller2017}. 

\section{Experimental setup and methods}

The experiments are performed on GaAs epitaxial QDs obtained by droplet etching and overgrowth, embedded in an Al$_{0.4}$Ga$_{0.6}$As matrix at $4.2$\,K ~\cite{Huo2013}. The photoluminescence from the ensemble is centered around $780$\,nm. QDs in the ensemble can be brought into resonance with the Rb D2 line using strain tuning as detailed in Ref.~\cite{Jahn2015}. This is a powerful feature. However, the spin properties of these QDs are presently unexplored. In particular, spin-pumping has not previously been achieved on these QDs.

The QDs can be charged with an excess hole or electron by illuminating the sample with weak, nonresonant laser light at $633$\,nm~\cite{Nguyen2012}. Here, we study the line identified as the positively charged exciton $X^{1+}$ of one single QD. The identification is based on the widely different $g$-factors of electrons and holes in GaAs~\cite{Ulhaq2016}; the electron $g$-factor is assumed to be negative. The QD is subjected to a magnetic field of 2.8\,T along the sample growth axis and parallel to the optical axis (Faraday geometry) resulting in a pair of spin-preserving optical transitions. For the chosen QD, the electron and hole g-factors are determined to be $g_e = -(0.05 \pm 0.01)$ and $g_h = (0.41 \pm 0.02)$ based on the energy splittings of the four transitions. The two spin-preserving transitions are separated in frequency by $\approx 18$ GHz (Fig.~\ref{Fig1}(a)). 

Figure~\ref{Fig1}(b) shows the polarization-based dark-field microscope used to collect the resonance fluorescence on resonant excitation~\cite{Kuhlmann2013a}. Linearly-polarized laser light propagates in a single mode through an excitation port, and the orthogonally-polarized light scattered by the QD is collected at a separate detection port. A polarizing beam splitter separates the scattered light from the excitation. Exquisite fine control of the polarization suppresses back-scattered laser light at the detection port up to 80 dB, and we observe resonance fluorescence (RF) with a signal-to-background ratio up to $100:1$. A ZrO$_2$ solid-immersion lens mounted onto our sample in combination with an aspheric lens of numerical aperture of 0.77 enhances the collection efficiency. 

Electro-optic intensity modulators (EOM, Jenoptik, 200 ps rise time) driven by a fast arbitrary waveform generator (AWG, Tektronik 7122C) allow excitation pulses with tailored intensity profiles to be generated. The QD output is coupled into a fiber and guided either to a spectrometer equipped with electron multiplying charge coupled device or to single photon detectors (APDs) connected to a time-correlated single-photon counting module (Picoharp 300). The photons' temporal profiles are reconstructed with a resolution of $512$\,ps by recording histograms of APD detection events. 

To study the spectral properties of the QD photons, we added a Fabry-P\'{e}rot etalon (FP) to the detection arm (12.9\,GHz free spectral range, 250\,MHz linewidth). The FP is frequency tuned via a heater, and the temperature is feedback-controlled; the FP has high long-term stability. A spectrum is obtained by recording the number of detected photons after the FP etalon during 100\,s as a function of the etalon detuning $\Delta_{\rm FP}$.

\section{QD spin dynamics}

\begin{figure*}[tb]
\includegraphics[width=150mm]{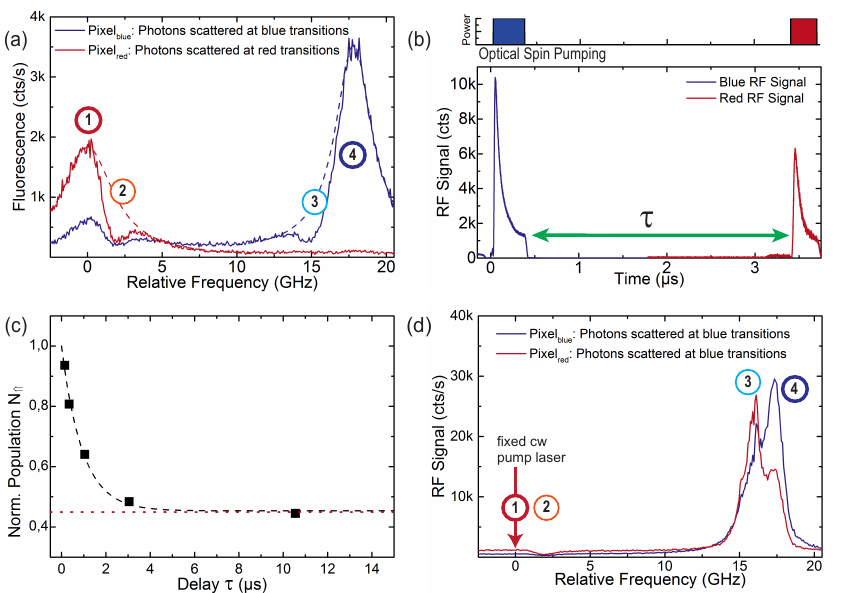}
\caption{Observation of optical spin pumping. (a) $X^{1+}$ resonance fluorescence (RF) spectrum in the Faraday configuration at $B_z=2.8$\,T. The CCD-spectrometer ($9$\,GHz resolution) partially resolves light scattered on the red transitions \raisebox{.5pt}{\textcircled{\raisebox{-.9pt} {1}}} and \raisebox{.5pt}{\textcircled{\raisebox{-.9pt} {2}}} (red trace), and the blue transitions \raisebox{.5pt}{\textcircled{\raisebox{-.9pt} {3}}} and \raisebox{.5pt}{\textcircled{\raisebox{-.9pt} {4}}} (blue trace) on two distinct pixels, with residual leak of the red signals counted on the ``blue" pixel. The dip in the red (blue) trace compared to the fitted lorentzian profile (dashed line) shows the enhancement of the optical spin pumping that depopulates the ground state $\ket{\Uparrow}_z$ ($\ket{\Downarrow}_z$) when the driving laser is resonant with the spin-flipping transition \raisebox{.5pt}{\textcircled{\raisebox{-.9pt} {2}}} (\raisebox{.5pt}{\textcircled{\raisebox{-.9pt} {3}}}). (b) Time-resolved fluorescence observed under pulsed resonant excitation, alternately pumping the transition \raisebox{.5pt}{\textcircled{\raisebox{-.9pt} {4}}} and \raisebox{.5pt}{\textcircled{\raisebox{-.9pt} {1}}} at saturation with a delay $\tau$. The exponential decays result from optical spin pumping that sequentially prepares $\ket{\Uparrow}_z$ and $\ket{\Downarrow}_z$ with time constant $\tau_{\rm opt} = 50$\,ns. (c) Spin relaxation dynamics. The exponential fit (dashed line) gives an effective $1/e$ spin thermalization time of $0.95\,\mu$s. Dotted line shows Boltzmann equilibrium. (d) Same as (a) but with additional CW laser (1) driving the spin-preserving transition \raisebox{.5pt}{\textcircled{\raisebox{-.9pt} {1}}} at saturation. The spin remains optically pumped in $\ket{\Downarrow}_z$ except when the scanning laser is resonant with the spin-preserving transition \raisebox{.5pt}{\textcircled{\raisebox{-.9pt} {4}}} (photon scattering at transitions \raisebox{.5pt}{\textcircled{\raisebox{-.9pt} {1}}} and \raisebox{.5pt}{\textcircled{\raisebox{-.9pt} {4}}}) or the spin-flipping transition \raisebox{.5pt}{\textcircled{\raisebox{-.9pt} {3}}} (photon scattering mostly at transition \raisebox{.5pt}{\textcircled{\raisebox{-.9pt} {1}}}, with residual leak of red fluorescence counted on the blue pixel).}
\label{Fig2}
\end{figure*}

First, we demonstrate the optical initialization of the QD hole spin in the Faraday geometry. We work initially at moderate nonresonant (633 nm) laser intensities ($0.15$\,nW/$\mu$m$^2$). The resonance fluorescence (RF) spectrum of $X^{1+}$ is shown in Fig.~\ref{Fig2}(a). The red (blue) curve displays the rate of QD photons detected on the spectrometer around the spin-preserving transition frequency \raisebox{.5pt}{\textcircled{\raisebox{-.9pt} {1}}} (\raisebox{.5pt}{\textcircled{\raisebox{-.9pt} {4}}}) as we scan the frequency of a continuous wave (CW) excitation laser above saturation. The red (blue) RF peak is well fitted by a (power-broadened) lorentzian profile, except for a dip observed when the scanning laser is resonant with the spin-flipping transition \raisebox{.5pt}{\textcircled{\raisebox{-.9pt} {2}}} (\raisebox{.5pt}{\textcircled{\raisebox{-.9pt} {3}}}). Qualitatively, such dips in the RF signals show the enhancement of the optical spin pumping in which the spin, initially in a statistical mixture of the two spin states, is driven into one of the spin states.

To access the spin pumping and relaxation dynamics, we implement an all-optical method similar to Ref.~\cite{Lu2010} based on time-resolved resonance fluorescence (TRRF) measurements. The two-color excitation sequence is illustrated in Fig.~\ref{Fig2}(b). Acousto-optic modulators are used to create pulses from two CW lasers, resonant with the spin-preserving transitions \raisebox{.5pt}{\textcircled{\raisebox{-.9pt} {1}}} and \raisebox{.5pt}{\textcircled{\raisebox{-.9pt} {4}}} respectively, which alternately pump the spin into $\ket{\Downarrow}_z$ and $\ket{\Uparrow}_z$. Figure~\ref{Fig2}(b) displays the TRRF signals when the pulses drive the spin-preserving transitions well-above saturation. Exponential fits (not shown) indicate optical spin pumping times $\tau_{\rm opt} = 50$\,ns ($\sim 2/\gamma$). The spin pumping time is much larger than the radiative emission time, $330$\,ps for this QD. This represents an experimental demonstration that the branching ratio is highly asymmetric $ \gamma / (\gamma + \Gamma) \sim 1:75$.

In a next step, the spin relaxation dynamics are investigated by increasing the delay $\tau$ between the two resonant pulses. Just after a $400$\,ns pulse on transition \raisebox{.5pt}{\textcircled{\raisebox{-.9pt} {4}}}, the spin is initialized in ground state $\ket{\Uparrow}_z$. Without laser excitation, the spin flips from $\ket{\Uparrow}_z$ to $\ket{\Downarrow}_z$ ($\ket{\Downarrow}_z$ to $\ket{\Uparrow}_z$) at a rate $\gamma_{\Uparrow \Downarrow}$ ($\gamma_{\Downarrow \Uparrow}$) due to interaction with its environment. When the next pulse on transition \raisebox{.5pt}{\textcircled{\raisebox{-.9pt} {1}}} arrives, the RF signal amplitude is proportional to the spin population left in $\ket{\Uparrow}_z$. Figure~\ref{Fig2}(c) shows the decay of the population $N_{\Uparrow}$ as the delay $\tau$ increases. By solving rate equations, the populations ($N_{\Uparrow},N_{\Downarrow}$) both relax to Boltzmann equilibrium $ N_{\Uparrow}/N_{\Downarrow}=\gamma_{\Downarrow \Uparrow} / \gamma_{\Uparrow \Downarrow} = \exp(-g_{h} \mu_B B / k_B T)$ at an effective rate $\gamma_{\rm eff} =\gamma_{\Uparrow \Downarrow} + \gamma_{\Downarrow \Uparrow}$. By fitting the decay of $N_{\Uparrow}$ with an exponential, we extract an effective spin relaxation time $\gamma_{\rm eff}^{-1}$ of $0.95\,\mu$s which corresponds to a spin lifetime $\gamma_{\Uparrow \Downarrow}^{-1}$ of $1.75\,\mu$s at 2.8\,T. The optical spin pumping is thus much faster than the spin relaxation dynamics ($\tau_{\rm opt}^{-1} / \gamma_{\Uparrow \Downarrow}  \sim 35 $) which enables fast and efficient spin ground state preparation. 

Finally, the ability to drive the weak cross transitions is demonstrated in Figure~\ref{Fig2}(d). A CW laser (1) resonantly drives the red spin-preserving transition \raisebox{.5pt}{\textcircled{\raisebox{-.9pt} {1}}} at saturation, and a RF spectrum is recorded as the frequency of a second CW scanning laser (2) is tuned across the optical transitions. In this experiment, the red RF signal is almost constant at 1.2\,kcts/s over the scanning range, as laser (1) keeps on driving transition \raisebox{.5pt}{\textcircled{\raisebox{-.9pt} {1}}}. However, it reduces when the scanning laser (2) becomes resonant with the spin-flipping transition \raisebox{.5pt}{\textcircled{\raisebox{-.9pt} {2}}}. Moreover, a new peak is clearly observed when the scanning laser (2) comes into resonance with the spin-flipping transition \raisebox{.5pt}{\textcircled{\raisebox{-.9pt} {3}}}. Our interpretation is that the spin-flipping transitions are not just decay paths but can be driven in the Faraday geometry. Tuned to resonance \raisebox{.5pt}{\textcircled{\raisebox{-.9pt} {2}}}, laser (2) enhances the spin pumping achieved with laser (1). Tuned to resonance \raisebox{.5pt}{\textcircled{\raisebox{-.9pt} {3}}}, laser (2) disrupts the spin pumping achieved with laser (1). The ratio of the red RF signals when spin pumping is disrupted or present gives a spin preparation efficiency of $95$\,\%. 

We note that in our sample, the nonresonant excitation not only allows the QDs to be charged, but also controls the hole spin relaxation rate, and thus the spin preparation efficiency. This represents useful {\em in situ} control. For instance, the basic spectroscopy to establish the frequencies of the transitions can be conducted at high ($>30$\,nW/$\mu$m$^2$) nonresonant excitation (suppressed spin pumping, large RF signals); photon shaping is then implemented at low ($0.03$\,nW/$\mu$m$^2$) nonresonant excitation (high efficiency spin pumping $\geq 95$\,\%).

These experiments establish all the features required for generating single photons with a Raman process, namely spin initialization via optical pumping and a ``diagonal" spin-flipping transition which can be driven in the Faraday geometry.

\section{Raman single-photon pulse shaping}

\begin{figure*}[t!]
\includegraphics[width=150mm]{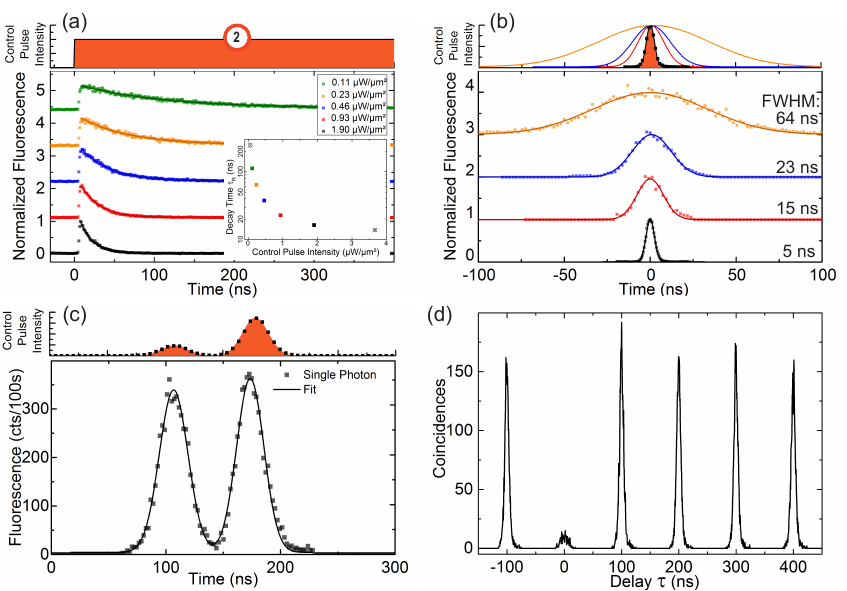}
\caption{Single-photon pulse shaping. (a) Exponential photon waveforms obtained with a square control pulse. The single photon waveforms are shown (with an offset for visibility) as the intensity of the control pulse decreases. Inset: Tuning of the Raman photon duration $\tau_R$ with control pulse intensity. (b) Gaussian photon waveforms (gaussian control pulses) with FWHM duration of 5, 15, 23 and 64\,ns respectively. (c) Double gaussian photon waveform. Each curve from (a) to (c) corresponds to 10\,min integration, and 2\,ns time resolution. (d) Intensity autocorrelation of Raman photons with gaussian waveform (${\rm FWHM}=5$\,ns, $10$\,MHz repetition rate, 11\,h acquisition).}
\label{Fig3}
\end{figure*}

We demonstrate the pulsed generation of single Raman photons with tailored waveforms. We use a two-color excitation sequence similar to Fig.~\ref{Fig2}(b), addressing the transitions \raisebox{.5pt}{\textcircled{\raisebox{-.9pt} {4}}} and \raisebox{.5pt}{\textcircled{\raisebox{-.9pt} {2}}}. In a first step, the spin is prepared in $\ket{\Uparrow}_z$ using a pump pulse ($50-200$\,ns) on resonance with transition \raisebox{.5pt}{\textcircled{\raisebox{-.9pt} {4}}}. Subsequently, a second control pulse with frequency $\nu_L$ drives the spin-flipping transition \raisebox{.5pt}{\textcircled{\raisebox{-.9pt} {2}}} at a detuning $\Delta_L=\nu_L-\nu_2$. The sequence is repeated at a rate up to $10$\,MHz. The concept is to induce a single spin-flip along with the emission of a single blue Raman photon. (The reverse scheme starting in $\ket{\Downarrow}_z$ and emitting a red Raman photon by driving the weak transition \raisebox{.5pt}{\textcircled{\raisebox{-.9pt} {3}}} is an equivalent concept). By adjusting the temporal envelope of the control pulse, a user-defined temporal structure is imprinted on the Raman photon waveform.

Figures~\ref{Fig3}(a-c) show photon waveforms obtained for different control pulses shapes close to resonance ($\Delta_L=0$). With square control pulses (Fig.~\ref{Fig3}(a)), the quantum dot output exhibits an abrupt onset (limited by the rise time of the EOM) followed by an exponential decrease in the trailing edge with time constant $\tau_R$. By decreasing the control pulse power, we can adjust the duration $\tau_R$ of the single photons from $14$\,ns to $245$\,ns, which is respectively about two and three orders of magnitude longer than the intrinsic radiative lifetime of the trion states ($330$\,ps). With the perspective of optimizing the interface of our single-photon source with a rubidium quantum memory~\cite{Gorshkov2008}, we also demonstrate the ability to tailor the temporal envelopes of the single photon wave packets. An efficient starting point for memory optimization is to use gaussian profiles of chosen duration. Using gaussian control pulses, we generate gaussian single photons of full-width-at-half-maximum (FWHM) duration ranging from 5\,ns to 64\,ns (Fig.~\ref{Fig3}(b)). 
As a final example, a more complex pulse shape, we address the possibility of splitting a single photon over two distinct time bins. Such photons have application in robust long-distance quantum communication protocols~\cite{Brendel1999,Marcikic2002,Humphreys2013,Jayakumar2014}. To do this, we apply a double gaussian waveform to the control laser. The quantum dot ouput mimics the control (Fig.~\ref{Fig3}(c)). 

To confirm the single-photon nature of the Raman light stream, we measured the second-order coherence for each temporal waveform using a standard Hanbury Brown-Twiss (HBT) setup. The two APDs were gated such that only photons emitted during the Raman generation phase (and not the initialization phase) were counted. The observed coincidences form a series of spikes separated by the sequence period, each with a shape related to the photon wave packet. A nearly vanishing peak at zero delay demonstrates that at most one single Raman photon is emitted during one sequence. The raw multi-photon emission probability ($g^{2}(0)$) is extracted by computing the ratio of coincidence events of the central peak to the mean of the next five neighboring peaks. Figure~\ref{Fig3}(d) shows the intensity correlation histogram obtained for the 5\,ns gaussian photons with $g^{2}(0)=0.12$. Similar values were obtained for all the different waveforms: 0.10-0.15 for exponentials, 0.12-0.33 for gaussians and 0.26 for the double gaussian. Residual coincidences originate from the detection of photons off-resonantly scattered by the QD either on transition \raisebox{.5pt}{\textcircled{\raisebox{-.9pt} {1}}} before the Raman flip, or on transition \raisebox{.5pt}{\textcircled{\raisebox{-.9pt} {4}}} after the emission of the first Raman photon, i.e. after the spin has flipped to $\ket{\Downarrow}_z$. No selection rules in the Faraday configuration prohibit the linearly polarized control laser from driving off-resonantly the transitions \raisebox{.5pt}{\textcircled{\raisebox{-.9pt} {1}}} and \raisebox{.5pt}{\textcircled{\raisebox{-.9pt} {4}}}. In practice, we find that the control laser intensity and detuning can be adjusted to reach a good compromise between high single-photon generation rate and low multi-photon emission probability.

\section{Spectral properties of Raman photons}

\begin{figure*}[tb]
\includegraphics[width=150mm]{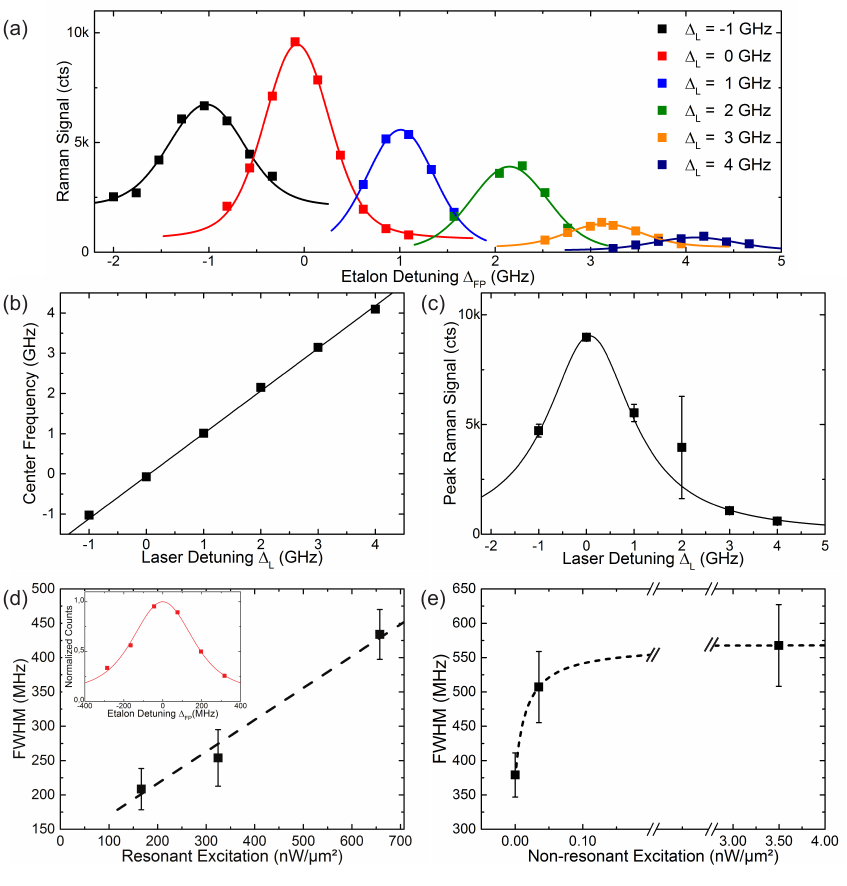}
\caption{Spectral properties of the Raman photons. (a) Spectra of 50\,ns gaussian single photons measured for different control laser detunings $\Delta_L$ (2\,MHz repetition rate, 100\,s integration). Voigt fits (solid lines) are used to extract values and error bars for the center frequency (b), peak (c) and linewidth of the Raman light stream deconvoluted from the (lorentzian) etalon transmission profile. (d) Increase of the spectral linewidth with control pulse intensity for a fixed temporal profile with the minimum nonresonant intensity (the dashed line indicates a linear fit). Inset: Narrowest spectrum with deconvoluted linewidth (FWHM) of 200\,MHz. (e) Increase of the spectral linewidth with nonresonant intensity. The saturation fit (dashed line) is a guide to the eye.}
\label{Fig4}
\end{figure*}

Besides tunability and purity, quantum protocols based on two-photon interference require sources of single photons with a high degree of indistinguishability. By comparing the spectral linewidth and the Fourier transform of a given temporal wave packet, one can infer the degree of indistinguishability that would be measured in a Hong-Ou-Mandel interference experiment. This is a very stringent test: it compares photons generated at widely different times. 

Figure~\ref{Fig4}(a) shows the spectra corresponding to 50\,ns gaussian single photons obtained for different excitation detunings $\Delta_L$ from the $\ket{\Downarrow}_z \rightarrow \ket{\Downarrow \Uparrow \uparrow}_z$ resonance. These measurements were performed with a moderate nonresonant laser intensity of $0.15$\,nW/$\mu$m$^2$ to increase signal count rates. Each curve is fitted by a Voigt profile to extract the center frequency, amplitude and linewidth, after deconvolution from the etalon transmission profile. A free offset allows an estimation of the number of unwanted (background) counts due to photons scattered before or after the Raman flip. As expected, the center frequency of the Raman signals shifts linearly with the laser detuning $\Delta_L$, while its peak amplitude follows a lorentzian profile in $\Delta_L$, as shown in Fig.~\ref{Fig4}(b) and (c), respectively. However, the expected decrease of the linewidth with $\Delta_L$ at large values is not observed; instead, it retains a value of $\sim\,700$\,MHz. 

To understand this, we studied the influence of laser intensities on the Raman photon linewidths from both the 780\,nm resonant control and the 633\,nm nonresonant laser. Initially, the nonresonant intensity was set to a low level ($\sim 0.01$\,nW/$\mu$W$^2$) to minimize charge noise in the QD environment, and we measured the variations of the spectral linewidths as we increased the intensity of a square resonant control pulse. From the data shown in Fig.~\ref{Fig4}(d), the linewidth increases linearly with the intensity of the control pulse. This points to a broadening mechanism involving laser-induced mixing between long- and short-lived states. Here indeed, due to the absence of strict polarization selection rules, the off-resonant couplings of the control field between $\ket{\Uparrow}_z$ and $\ket{\Downarrow \Uparrow \uparrow}_z$, and $\ket{\Downarrow}_z$ and $\ket{\Downarrow \Uparrow \downarrow}_z$ are expected to reduce the effective hole spin coherence. This effect is also responsible for the broadening of the spectrum observed in Fig.~\ref{Fig4}(a) when the laser comes close to resonance with the $\ket{\Uparrow}_z \rightarrow \ket{\Downarrow \Uparrow \uparrow}_z$ transition \raisebox{.5pt}{\textcircled{\raisebox{-.9pt} {1}}} at $\Delta_L = -1$\,GHz. However, at the cost of reduced single-photon emission efficiency, lower excitation intensity enables the generation of Raman photons with linewidths as low as 200\,MHz (see Fig.~\ref{Fig4}(d), inset), which is about an order of magnitude narrower than the 1.1\,GHz linewidth of the excited states measured in RF at low saturation.

Finally, we set the resonant square pulse to an intermediate peak intensity, and we measure the variations of the spectral linewidth with the intensity of the nonresonant laser. While the single-photon emission rate increases quickly as the QD becomes more active, we again observe a broadening of the linewidth up to about twice the initial value (Fig.~\ref{Fig4}(e)). We attribute this additional broadening to charge noise in the environment of the QD, which increases with the nonresonant intensity as more and more charges are optically excited in the Al$_{0.4}$Ga$_{0.6}$As matrix surrounding the QD. In our sample without charge control, fine tuning of the nonresonant power is thus required to reach a compromise between large single-photon emission rates and narrow emission linewidth.

\section{Conclusion and outlook}

In summary, we have demonstrated a fast single-photon source based on an epitaxial GaAs QD that generates on-demand single Raman photons with controlled temporal profiles at the wavelength of the rubidium D2 line. 

Reaching the Fourier-transform limit eventually requires to engineer the electronic and photonic QD environments. First, embedding the QDs in a pin-diode type structure would enable deterministic charge control, eliminating the need for an additional nonresonant excitation~\cite{Warburton2013}. Furthermore this capacitor-like structure is known to suppress charge noise. In the case of self-assembled InGaAs QDs, this results in close-to-transform-limited optical linewidths~\cite{Kuhlmann2013b,Kuhlmann2015}, and long $T_{2}^{*}$ times for the hole spin~\cite{Prechtel2016}. Secondly, adding a photonic structure to enhance the collection efficiency would enable operation at lower resonant power and larger detunings. This will improve the photons' properties, and could ultimately provide deterministic spin-photon entanglement using cavity-stimulated Raman spin-flip~\cite{Sweeney2014}. 

Even with the present performance, the demonstrated properties of our source make it immediately suitable for investigating EIT-based single-photon storage and retrieval in warm rubidium vapors~\cite{Wolters2017}. The ability to control the temporal profile of the photon wave packets opens the way for memory optimization using optimal control methods~\cite{Rakher2013}. Such a semiconductor-atom interface will form the basis for studies on hybrid entanglement between collective atomic spin-wave excitation and single semiconductor spins, as well as between distant atomic quantum memories in a quantum network.

\begin{acknowledgments}
We acknowledge financial support from NCCR QSIT and the European Union Seventh Framework Programme 209 (FP7/2007-2013) under Grant Agreement No. 601126 210 (HANAS). J.W. received funding from the Marie Skłodowska-Curie Actions of the EU Horizon 2020 Framework Programme under Grant No. 702304 (3-5-FIRST). J-P.J. and L.B. made equal contributions to this work. 
\end{acknowledgments}

\bibliographystyle{apsrev4-1}

%

\end{document}